# SRAM-based Physically Unclonable Function using Lightweight Hamming-Code Fuzzy Extractor for Energy Harvesting Beat Sensors

Hoang-Long Pham, Duy-Hieu Bui, Xuan-Tu Tran
VNU Information Technology Institute
Vietnam National University
Hanoi, Vietnam
Corresponding author's email: hieubd@vnu.edu.vn

Orazio Aiello
DITEN
University of Genoa
Genoa, Italy

*Abstract*—Batteryless energy harvesting IoT sensor nodes such as beat sensors can be deployed in millions without the need to replace batteries. They are ultra-low-power and cost-effective wireless sensor nodes without the maintenance cost and can work for 24 hours/365 days. However, they were not equipped with security mechanisms to protect user data. Data encryption and authentication can be used to secure beat sensor applications, but generating a secure cryptographic key is challenging. In this paper, we proposed an SRAM-based Physically Unclonable Function (PUF) combining a high-reliability bit selection algorithm with a lightweight error-correcting code to generate reliable secure keys for data encryption. The system employs a feature of beat sensors, in which the microcontroller is powered on to transmit the ID signals and then powered off. This fits the SRAM-based PUF requirement, which needs the SRAM to be powered off to read out its random values. The proposed system has been evaluated on STM32 Cortex M0+ microcontrollers and has been implemented to protect important data on beat sensors.

## I. INTRODUCTION

IoT sensor nodes are expected to be deployed in mass with thousands or millions of devices. They are ultra-low-power and low-cost wireless communication devices powered by batteries or energy harvesting. Batteryless energy harvesting IoT sensor nodes are preferable because they reduce the maintenance cost. Beat sensors are the devices following this trend. Beat sensors were first introduced in 2015 [1] with small sizes and the ability to harvest energy for their operations.

The basic operating principle of beat sensors is that the sensor nodes do not use ADC to read the sensor values and transmit them through wireless communication. Instead, beat sensor nodes only transmit the device's ID. The receiver receives this ID signal and computes the interval between two consecutive data transmissions from the sensor nodes. This time interval is proportional to the measured values at the sensor node[2]. The microcontroller and the RF module are often turned off, so beat sensors consume little energy and can be used with energy harvesting.

The efficiency of beat sensors leads to a security problem when any agent with an RF receiver with the proper configuration can capture the beat sensors' packet and calculate the interval between transmissions based on the ID signal. This exposes the measured value to the attacker. The attacker can replay the captured packet more dangerously, changing the interval time. Consequently, the attacker can change the measured value without detection. This security issue is critical in IoT systems and can not be ignored.

Beat sensors can be protected using state-of-the-art data encryption and security mechanisms. However, strong security also leads to high power and energy consumption, an obstacle to applying security to energy harvesting systems. The work in [3] proposed a lightweight data encryption mechanism with a small packet size to ensure security with a marginal increase in energy consumption. Lightweight data encryption can protect large beat sensor networks by encrypting the ID signal using Pre-Shared Key methods. However, the whole network is revealed if the key is leaked to the attackers. Therefore, beat sensors need a method to secure the Pre-Shared Key as proposed in[3].

Physically Unclonable Function (PUF) is a good candidate for further increasing the security of beat sensors. A secure PUF can generate a unique and random "fingerprint." It can be used to generate a cryptographic key to encrypt sensitive data, such as the Pre-Shared Key. Cryptographic keys must be stored in a beat sensor system in non-volatile memory, such as Flash or EPROM. Therefore, attackers can apply reverse engineering techniques to exploit these data. By using SRAM-based PUFs combined with secure debugging, the cryptographic key used to secure the communication in beat sensors [3] is not stored as clear text in memory but encrypted by a key generated from PUFs. The attackers cannot perform reverse engineering by simply reading non-volatile memory contents. In this work, we propose an SRAM-based PUF that can be used to generate security keys for encrypting sensitive data on beat sensors. In combination with secure storage and authenticated debugging, it will prevent the attackers from reverse engineering attacks. The proposed mechanism can generate the cryptographic key for each device to join the network in a system with a more power budget than beat sensors. This is critical in IoT applications that use LoRaWAN communication. The proposed method combines a bit selection algorithm with a lightweight error correction code to create a lightweight fuzzy extractor for PUFs. The proposed fuzzy extractor allows a bit flipped in the SRAM-based PUF's response. Instead of using complex methods that consume excessive hardware resources to build the SRAM-based PUF, we focus on the enrollment process by using a highly stable bit selection algorithm performed on the server to reduce the number of bits flipped in actual usage. As a result, a lightweight error correction code, Hamming code, can be used in the fuzzy extractor on the device to minimize energy consumption while still maintaining power/energy efficiencies.

The rest of this paper is organized as follows. Section II presents the related works on SRAM-based PUFs and fuzzy extractors. Section III discusses the fundamental principles of our proposed SRAM-based PUFs, fuzzy extractor, and bit selection algorithm. The experimental setup and evaluation results of the proposed method will be presented in Section IV. Finally, there are some conclusions and perspectives in Section V.

## II. RELATED WORKS

PUFs are hardware-based security primitives that can extract the physical characteristics of the components related to the variation in the manufacturing process to generate an unclonable "fingerprint" [4]. This characteristic distinguishes PUF from other hardware-based security methods because attackers cannot clone the device's intrinsic physical characteristics. PUF can be described as a function where different input sequences (Challenge $C$) applied to a PUF on different devices will produce different results (Response $R$). A challenge and its response form a challenge-response pair (CRP). Based on the number of CRPs, PUFs are divided into two main types: strong PUFs and weak PUFs. Strong PUFs have a large number of unique CRPs and are typically used in authentication protocols. In contrast, weak PUFs have only few CRPs, even only one, and can be used in secure key generation applications. PUFs provide an effective solution to IoT devices' security challenges with a high level of security without requiring excessive power/energy consumption.

SRAM memory is common in embedded systems. When SRAM is powered up, each memory cell in SRAM has a favored power-up state depending on manufacturing process variations. This variation in SRAM is unique and can be used to generate a physically unclonable fingerprint that is different across various ICs. Therefore, an SRAM-based PUF can be built by measuring the state of SRAM right after being powered on. SRAM-based PUFs are categorized as weak PUFs because it has only few CRPs. The addresses of memory cells can serve as challenges while their power-up values act as responses. The stability of SRAM PUF is significantly affected by factors such as the device's temperature, voltage, and chip lifespan.

The authors in [5] proposed a secure key generator by using SRAM-based PUF and a fuzzy Extractor and evaluated the bit error rate of the proposed method under different temperature conditions. They used BCH code as the error-correcting code to cope with the errors in the SRAM outputs. The proposed method achieved high accuracy but uses a complicated error correction code which is not suitable for beat sensors.

The work in [6] proposed an aging-injection-based method to address the uniformity and reliability issues of SRAM-based PUF. The proposed method utilizes an aging mechanism called negative-bias-temperature instability (NBTI) to make SRAM-based PUF more uniform and reliable.

The work in [7] proposed a secret-generation scheme with helper data for SRAM-based PUF. Error correction codes are an essential part of this scheme. This work uses a practical error correction construction for PUF-based secret generation based on polar codes. The resulting scheme can generate 128-bit keys using 1024 SRAM bits and 896 helper data bits and achieve a failure probability of $10^{-9}$ or lower for a practical SRAM PUF setting with a bit error probability of 15%.

The work in [8] uses almost the same mechanism as in [4], providing proof of concept for using SRAM-based PUFs to generate private keys for IoT devices. The author utilized a custom-made Arduino mega shield to extract the fingerprint from the SRAM chip on demand. They utilized the concepts of ternary states to exclude the cells that are easily prone to bit flips, allowing them to extract stable bits from the SRAM's contents. Using the custom-made software for their SRAM device, they can control the error rate of the PUF to achieve an adjustable memory-based PUF for key generation. They also utilized several fuzzy extractor techniques based on different error correction coding methods to generate secret keys from the SRAM PUF and study the trade-off between the false authentication rate and the false rejection rate of the PUF.

The authors in [9] used the bit selection methods to alleviate the noise of SRAM-based PUF's responses. Generally, this issue is solved by employing complex Error Correction Codes (ECCs). However, ECCs significantly increase the implementation cost, which is unsuitable for embedded systems. The complexity reduction in the SRAM-based PUF fuzzy extractor is important when applying SRAM-based PUF in IoT sensor nodes, especially for energy-harvested devices.

In[10], the authors analyzed the stability of SRAM elements and concluded that SRAM cells tend to be more stable when surrounded by other stable cells. The authors proposed evaluation criteria to select SRAM cells with good stability and proposed a bit selection algorithm for SRAM that can be used for PUF and security key generation applications.

In summary, current research has provided valuable knowledge and methods for generating secure keys based on SRAM-based PUFs for IoT devices. Based on these analyses, we proposed a key generation method that leverages the strengths identified in the aforementioned studies. Our proposed method focuses on a highly stable bit selection method that allows a one-bit error in the SRAM-PUF output. One-bit errors can be corrected in the fuzzy extractor using Hamming code. Our proposed method creates an efficient, cost-effective, energy-saving key generation scheme tested on STM32 Cortex M0+ energy harvesting beat sensors.

## III. PROPOSED METHOD

This paper presents our approach to generating security keys for the Beat sensor using SRAM-based PUFs, a highly stable bit selection algorithm, and a lightweight fuzzy extractor that allows one bit to be flipped in the PUF's responses. The proposed method focuses on the bit selection algorithm to enhance the efficiency of SRAM-based PUFs and reduce the computational complexity of the error correction code algorithm in the fuzzy extractor. Typically, multi-bit error correction code can be used to correct data with large deviations. However, if the input data deviation is small, the lightweight error correction code can be used to lower memory footprint and energy consumption. The following subsections detail our proposal.

### A. SRAM-based PUF enrollment and the proposed bit selection algorithm

The enrollment process is required to extract the characteristics of the devices before the PUF can be used. In

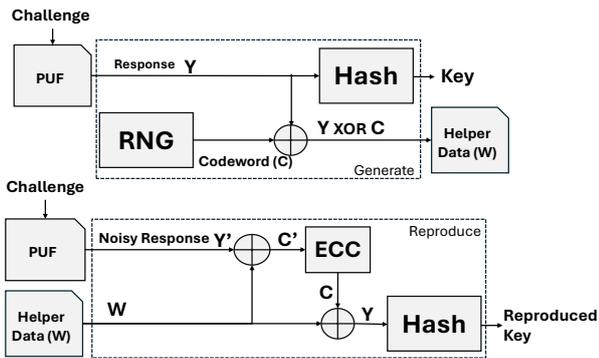

Fig. 1. Typical Fuzzy Extractor Mechanism.

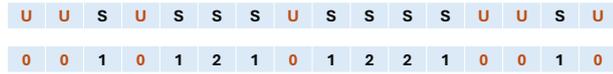

Fig. 2. Weight assignment.

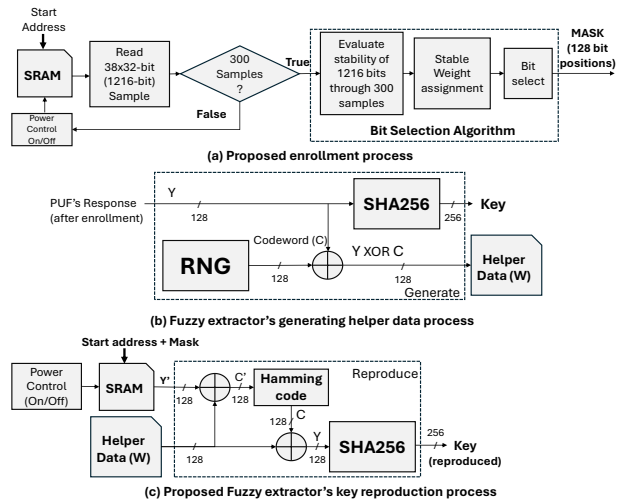

Fig. 3. Proposed Fuzzy Extractor combined with Bit Selection Algorithm.

**Algorithm 1:** Check stability

**Input**: A list of '**N sample Responses**'
**Output**: An array '**status of position**'
1: **procedure** Marking
2:    N: Total number of Responses
3:    **for** $i = 0$ to *Response length-1* **do**
4:       **for** $j = 1$ to N **do**
5:          **if** (Cell $i$ is stable across the N sample) **then**
6:             Mark cell $i$ = S
7:          **else**
8:             Mark cell $i$ = U
9:          **end if**
10:      **end for**
11:   **end for**

**Algorithm 2:** Weighting Bits Position [10]

**Input**: An array '**stable position**'
**Output**: An array of '**weights**' with the weight value of each stable position
1: **procedure** WeightingBit
2: **initialize array**: *weights*
3:    **for** $i = 0$ to '*length of stable position – 1*' **do**
4:       **if** ($i$ is the first/last index of '*stable position*') **then**
5:          weights[$i$] = 1
6:       **else**
7:          set a variable *weight* to 1
8:          **initialize** $j = 1$
9:          **while** $i - j >= 0$ and $i + j <$ *length of stable position* **do**
10:            **if** (stable position[$i - j$] < stable position[$i$] – $j$ or stable position[$i + j$] < stable position[$i$] + $j$)
11:               exit loop
12:            **else**
13:               *weight* += 1
14:               $j$ += 1
15:            **end if**
16:            weights[$i$] = *weight*
17:         **end while**
18:      **end if**
19:   **end for**

**Algorithm 3:** Select Position

**Input**: An array '**weights**' of size n, representing weights assigned to positions, threshold value '**T**'
**Output:** An array '**mask**' containing positions with weights greater than T
1: **procedure** SelectPositions(weights, T)
2: **initialize array**: *mask*
3:    l = 0
4:    **for** i = 0 to n - 1 do
5:       **if** weights[i] >= T **then**
6:          mask[l] = i
7:          l += 1
8:       **end if**
9:    **end for**

the case of an SRAM-based PUF, the enrollment process selects the stable bits and stores these locations in a database along with the helper data to perform error correction for fuzzy matching. Fuzzy matching in the fuzzy extractor allows a small variation in the PUF's output, as shown in Fig. 1. This is important because the PUF's output may vary depending on the working conditions. The stability of SRAM-based PUF is significantly affected by factors such as the device's temperature, voltage, and lifespan. In addition, the error correction code may leak information about the PUF's responses, which is not good for security. Therefore, optimizing the error correction code is important for SRAM-based PUF to reduce the complexity and power consumption [9]. In this paper, we implemented the bit selection method to reduce the bit error rate to one bit per PUF's response. The bit selection method can be done on the server during the enrollment phase, as shown in Fig. 3a. Therefore, the proposed solution will not affect the runtime and power consumption at the sensor node. The enrollment phase can be done as follows:

- Collect a large amount of data from a fixed SRAM address when the device has just powered up under various environmental temperature and voltage conditions. The length of each data collection is 1,216 bits, and at a fixed address, it will be collected N times.

- Examine each position in a block of 1,216 bits, then selecting the 128 positions that satisfy the stability requirement. These selected positions are called masks, which filter responses from the SRAM-based PUF to ensure the best stability for secure key generation.

However, each mask can only be used with a fixed initial SRAM address. When the base address changes, the enrollment process must be repeated. Different start addresses can be enrolled and stored in a database in the server so we can use different start addresses later on in the device. The enrollment process aims to achieve the highest efficiency and quality for SRAM-based PUF's responses for secure key generation.

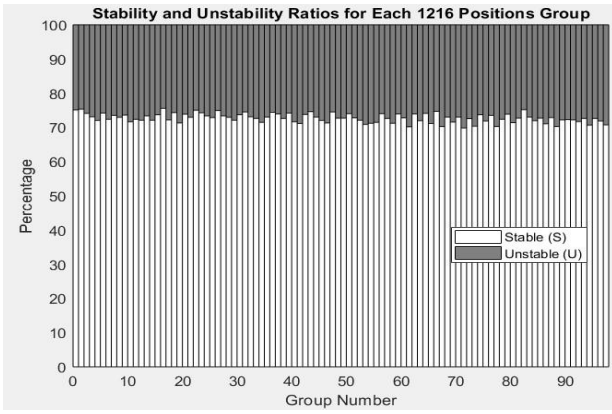

Fig. 5. Stability and Instability of 120,000 bits (3,750 address) SRAM, each block contains 1,216 bits (38 address of 32-bit data) at 25°C

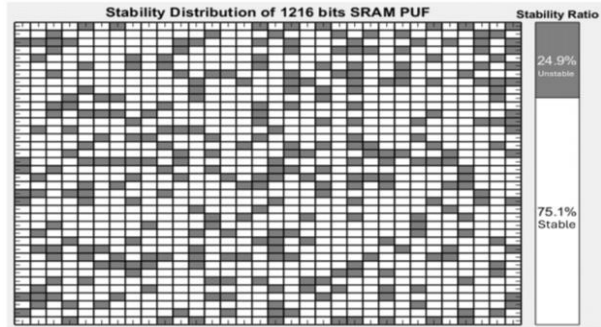

Fig. 6. Stability distribution of 1,216-bit block SRAM.

### B. Generate Cryptographic Keys from PUF using a Fuzzy Extractor.

Cryptographic key generation using PUF's responses needs randomness assessment and error correction code. The characteristics of entropy sources in the components that constitute PUFs, combined with factors such as environmental temperature and supply voltage, result in noisy PUF responses. As a result, it is difficult to reproduce the PUF responses accurately in the initial runs. PUFs' responses are often used as inputs to hash functions or key derivation functions to generate cryptographic keys. However, even a bit flip in the responses results in many bit errors in the hash function output or key derivation function. Moreover, in many cases, the response bit string generated by PUFs does not meet the criteria for randomness. Temperature and voltage variations tend to bias the bit string towards '0s' or '1s'. Secure sketches, strong extractors, and fuzzy extractors are functions designed to address these issues. Many error correction algorithms have been developed to correct errors in response strings. The most commonly used algorithm is to generate helper data as an additional source of information during the bit string generation process, which will later be used to aid in correcting bit-flipped errors during reconstruction. The helper data is stored publicly, so it must disclose as few as possible about the bit strings it is used to correct.

Fuzzy Extractors consist of two main procedures. The first one is the key generation procedure called "Generate" while the other is the key reproduction procedure called "Reproduce" as shown in Fig. 1. The PUF will generate a response string $Y$ based on the input challenge. A random codeword $C$ (independent of $Y$) is used to create an n-bit helper data string W as in (1).

$$W = Y \oplus C \quad (1)$$

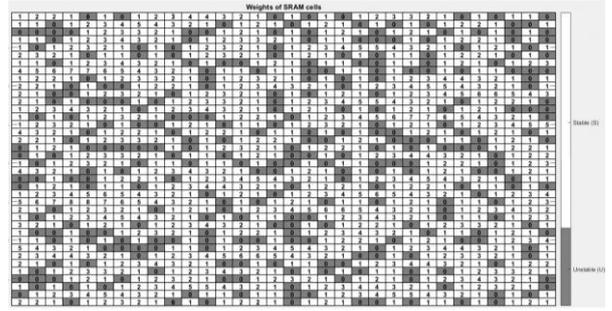

Fig. 4. Weight of SRAM cells in the 1,216-bit block.

TABLE I. AVERAGE NUMBER OF SELECTED BITS AT VARIOUS THRESHOLD VALUE T

| Threshold T | 1 | 2 | 3 | 4 | 5 |
|---|---|---|---|---|---|
| The average number of bits selected | 900.98 | 489.47 | 254.6 | 127.3 | 56.25 |

In subsequent iterations, the PUF will generate a different response $Y'$. Another random codeword $C'$ is generated as in (2).

$$C' = Y' \oplus W \quad (2)$$

Then, the codeword $C'$ will undergo error correction to obtain the corrected codeword $C''$ by (3).

$$C'' = Correct(C') \quad (3)$$

Using the obtained codeword $C''$, $Y''$ is calculated in (4).

$$Y'' = W \oplus C'' \quad (4)$$

If the number of different bits between $C$ and $C' < t$ ($t$ is the threshold of error correction code algorithm), then $C''$ is successfully corrected from $C'$ (meaning $C'' = C$), and in that case, $Y'' = Y$ (successfully reproduced). In this work, by using a highly stable bit selection algorithm, we can limit the value of $t$ to one, which results in using Hamming code to correct one error in the output of the proposed SRAM-based PUF.

### C. Proposed bit selection algorithm

In this subsection, we proposed a bit selection algorithm that selects the most stable bits, records their positions, and creates a mask to be used as a part of the challenge in SRAM-based PUFs. The mask will filter PUFs' output immediately after power-up to generate a highly stable response under various conditions. By using this mask, the response bits have high stability with a maximum of one error. As a result, lightweight error correction codes can be used to correct only one bit flip.

The authors in [10] analyzed and evaluated SRAM's outputs under various conditions that affect the stability of SRAM cells, including the correlations between the spacing of stable bits, meaning the most stable cells tend to be surrounded by other stable cells. The algorithms we employed in this work are based on the algorithm in [10], which takes into account the spatial correlations. In Algorithm 1, we gather a set of $N$ samples (SRAM data after being powered on) and assess them for stability. The length of each raw response in our work is 120,000 bits, corresponding to 120,000 positions. These positions are evaluated across $N$ Responses obtained earlier. Positions where the bits remain unchanged are marked as $S$ (Stable), while positions where the bits flip are marked as $U$ (Unstable), as shown in Fig. 2.

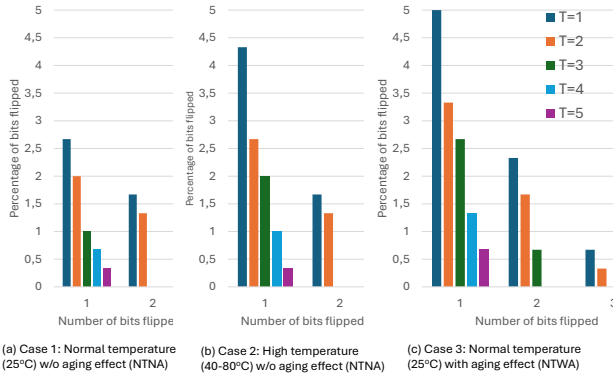

(a) Case 1: Normal temperature (25°C) w/o aging effect (NTNA)
(b) Case 2: High temperature (40-80°C) w/o aging effect (NTNA)
(c) Case 3: Normal temperature (25°C) with aging effect (NTWA)

Fig. 7. The percentage of occurrence in different test cases with various thresholds T.

We observe that the *S* positions often appear in contiguous clusters. Each cluster has a length of either *2n* or *2n+1* bits. A stable cell surrounded by stable cells tends to be more stable. After assessing the stability of the bits, we assign weights to the positions of the bits. In a cluster of *2n* stable bits, the two bits in the middle of the cluster at the positions *n* and *n+1* weigh *n*. The weights decrease from the middle of a cluster toward the two ends with the first and last bits in the cluster weight 1. Similarly, in a cluster of *2n+1* stable bits, the bit at position *n+1* weights *n+1*. The weights at the other positions in a cluster decrease gradually toward two ends. The entire process is presented in Algorithm 2.

In Algorithm 3, we evaluate the stability of bit positions in the sequences of obtained bit responses. Based on the weights of the stable bit positions, we select the positions with weights greater than or equal to a threshold *T*. This is to ensure that the selected bits have enough stable neighboring bits.

### D. Generate a key using SRAM PUF's Response.

In this work, we utilize Hammiyng code to reduce the complexity of error correction algorithms in the fuzzy extractor. The Hamming code is lightweight and easy to implement; however, it can only correct one error. Therefore, the proposed method selects stable bits to enhance the confidence in the PUF response and ensure no more than one error occurs. Thus, combining a highly stable bit selection algorithm with Hamming code ensures the reliability of the system while maintaining the power/energy budget on energy harvesting devices. Our method is shown in Fig. 3.

Fig. 3a shows the enrollment process where the bit selection algorithm examines the bit stability and generates a mask of the most stable position in the SRAM memory at a specific segment. The system is powered on and off multiple times to evaluate the stability of the SRAM cells, up to 300 times in this work. After that, the helper data will be generated as in Fig. 3b. Finally, the key can be generated on the sensor node as in Fig. 3c. Right after being powered on, the SRAM contents at the start address will be read and filtered using the mask generated in the enrollment process. Then, the response will be XORed with the helper data before being corrected using Hamming code. After that, the output is sent to the hash function to generate the secret key. Here, we use the SHA-256 algorithm, and the length of the resulting key is 256 bits, which is split into two keys for use.

## IV. EXPERIMENTAL RESULTS

### A. Experiment setup

This section implements and evaluates our proposed methods practically using multiple STM32 Nucleo L073RZ development boards, which use the same microprocessors and embedded SRAM as in beat sensors. The integrated SRAM on the STM32 M0+ microcontroller is employed to construct an SRAM-based PUF. The microcontroller on the STM32 development boards reads and transmits data from SRAM to the PC via UART for enrollment. The same procedure is done on beat sensors, but the data are sent through LoRa communication because beat sensors do not have UART communication to save energy consumption.

In our experiment setup, we conduct experiments to evaluate the responses obtained under different temperature conditions to simulate the environment temperature variation. The device is only powered on, then powered off entirely, as in beat sensors. We collect 300 consecutive samples every 5 seconds under different temperature conditions, each time gathering 3,750 addresses, each address with 32 bits (120,000 bits per sample) of the SRAM's initial values.

### B. Reliability of SRAM

In this experiment, we initialize an SRAM array to collect 300 samples of the initial values of this SRAM array from address 0x20001050 to address 0x20001EF6 (3,750 addresses), with each address containing 32 bits of SRAM. The first sampling condition is at room temperature, 25°C. The second sampling condition is when the board is heated to around 40-80°C. Each sample includes 120,000 bits, with each bit corresponding to one position, and these positions are evaluated over 300 samples. Positions where the initial value of the SRAM cell does not change throughout 300 samples are marked as *S*, and positions where changes are marked as *U*. We split these 120,000 positions into 1,216-bit blocks. Their statistics are shown in Fig. 5. It can be seen that the proportion of stable and unstable bits in each block does not differ significantly, remaining within the range of 72%-78%.

We randomly evaluate a 1,216-bit block, and the distribution of *S* (in white) and *U* (in grey) positions is shown in Fig. 6. It can be seen that the ratio of the S positions to the total number of positions in the 1,216-bit block is similar to the ratio of the entire *S* position to the total number of 120,000 bits.

### C. Reliability of SRAM-based PUF responses after using the bit selection algorithm

In this experiment, we evaluate the reliability of the response after it is selected using the bit selection algorithm. As previously mentioned, we split 120,000 bits of SRAM into blocks, each containing 1,216 bits. Each block is assessed for stability, and stable positions are weighted, as shown in Fig. 4. A minimum threshold of weighted values is used to filter out positions that meet the quality criteria.

Next, we analyze the impact of selecting the threshold value *T* under the influence of temperature and aging effects. Based on the analysis from the article[10], cells tend to be more stable when surrounded by other stable cells. We evaluate various threshold values. 98 of 1,216-bit blocks out of 120,000 sampled bits are evaluated to report the flipped bits under three conditions: normal temperature (25°C) without aging effect (NTNA) in Fig. 7a; high temperature (40-80°C) without aging effect (HTNA) in Fig. 7b; normal temperature (25°C) with 5 hours of aging effect (NTWA) in Fig. 7c. Under the condition with aging, the devices are powered on and operated continuously for 5 hours before the SRAM values are sampled. For each threshold under each condition, we take 300 continuous samples every 5 seconds.

From the obtained result, with the threshold values *T=1* and *T=2*, there is a rate of 2 bits flipping under all conditions. With threshold values *T=3*, there is a 2-bit flip under the conditions of 5 hours of aging effect, while with threshold values *T=4* and *T=5*, a maximum of 1-bit flips under all conditions on STM32 M0+ microcontroller.

Table I shows the number of bits selected for each threshold value, indicating that a threshold value of *T=5* yields the best results in this experiment. However, the number of highly stable bits that can be extracted should also be considered. If we want to choose additional bits to reach the lengths of 128 bits or 256 bits for security applications, the initial 1,216-bit block cannot be used. This would increase the number of necessary SRAM resource usage for SRAM-based PUFs. Threshold values of *T=3* and *T=4* have suitable lengths for security applications. However, the aging effect still significantly affects the threshold value of T=3, potentially causing two bits to flip.

This paper aims to use a lightweight error correction code for the device, specifically the Hamming error correction code, which can correct a maximum of one error. Therefore, the threshold value of *T=4* is appropriate. It ensures the reliability of the SRAM-based PUFs' response and the bit length of the response for key generation applications.

Table II compares our proposed method with the other works. Our method achieves a bit-flipped rate of 1% under NTNA conditions, 1.33% under HTNA conditions, and 1.667% under NTWA, which are significantly lower than the other works. This low error rate ensures reliability and stability in our applications. Our work uses Hamming code with low computational overhead compared to BCH or Polar codes used in the other works.

## V. CONCLUSIONS

In this paper, we presented a method for generating security keys for SRAM-based PUF using a highly stable bit selection algorithm combined with a lightweight fuzzy extractor to enhance the security of the beat sensor system. The bit selection algorithm is based on analyzing the stability of bits obtained from SRAM, considering the stability of neighboring cells. Based on the special operating principle of beat sensors, which involves powering the system for a short period and then cutting off power to the entire system, the stability of the positions selected by the bit selection algorithm is ensured and unaffected by aging effects with the result being a maximum of 1-bit flip under each evaluation condition. Due to the efficiency of the bit selection algorithm, the Hamming code can be used as a lightweight error correction code to generate secret keys to secure beat sensors.

ACKNOWLEDGMENT

This work has received funding from the European Union's Horizon Europe research and innovation program under the Marie SkłodowskaCurie grant agreement No 101086359.

TABLE II. RELIABILITY COMPARE TO OTHER WORKS

| Criteria | This work | [5] | [7] | [8] | [9] | [10] |
|---|---|---|---|---|---|---|
| Number of samples evaluated | **300 each condition** | 100/each condition | N/A | 1000 | 2000 for enrollment 200 for test | 100 fresh 200 burn-in |
| Bit Flipped Rate in Enrollment phase | **24.9%** | 4-10% | 15-25% | 54% | 52.21% | 22% NTNA 35% HTNA |
| Bit Flipped Rate in Normal Tempurature No Aging (NTNA) | **1% (1 bit flip)** | 0% | Failure of Polar code based is $3.47 * 10^{-10}$ - $10^{-6}$ with an error rate of 13-15% | 2-15% | 0.12-2.27% | 0.1%-1% 80 samples |
| Bit Flipped Rate in High Tempurature No Aging (HTNA) | **1.33% (1 bit flip)** | 6.5% | | N/A | 0-0.37% | N/A |
| Bit Flipped Rate in Normal Tempurature With Aging (NTWA) | **1.667% (1 bit flip)** | N/A | | N/A | 0.85-4.65% (after) 0-1.2227% (21 days) | 0.1-1.5% 80 samples |
| ECC | **Hamming code** | BCH code | Polar code | Polar code | N/A | N/A |